\documentclass[reprint,superscriptaddress]{revtex4-1}
\pdfoutput = 1
\usepackage{graphicx}
\usepackage{amsmath}
\usepackage{amssymb}
\usepackage{geometry}

\usepackage{lipsum}

\makeatletter
\renewcommand\@make@capt@title[2]{%
 \@ifx@empty\float@link{\@firstofone}{\expandafter\href\expandafter{\float@link}}%
  {\textbf{#1}}\@caption@fignum@sep#2\quad
}%
\makeatother

\begin{document}

\title{Nonlinear microwave photon-occupancy of a driven resonator strongly coupled to a transmon qubit}

\author{B. Suri} \email{surib@chalmers.se}\altaffiliation{Present address: Department of Microtechnology and Nanoscience, Chalmers University of Technology, 41296-Gothenburg, Sweden.}\affiliation{Laboratory for Physical Sciences, College Park, MD 20740, USA} \affiliation{Department of Physics, University of Maryland, College Park, MD 20742, USA}

\author{ Z. K. Keane}
\affiliation{Northrop Grumman Electronic Systems, Linthicum, MD 21090, USA}
\author{ Lev S. Bishop}
\affiliation{IBM T.J. Watson Research Center, Yorktown
Heights, NY 10598, USA}

\author{S. Novikov}
\affiliation{Laboratory for Physical Sciences, College Park, MD 20740, USA}\affiliation{Department of Physics, University of Maryland, College Park, MD 20742, USA}

\author{ F. C. Wellstood}
\affiliation{Department of Physics, University of Maryland, College Park, MD 20742, USA}
\affiliation{Joint Quantum Institute, University of Maryland, College Park, MD 20742, USA}
\affiliation{Center for Nanophysics and Advanced Materials, University of Maryland, College Park, MD 20742, USA}

 \author{B. S. Palmer}
 \affiliation{Laboratory for Physical Sciences, College Park, MD 20740, USA}
 \affiliation{Department of Physics, University of Maryland, College Park, MD 20742, USA}

\begin{abstract}
\noindent
We measure photon-occupancy in a thin-film superconducting lumped
 element resonator coupled to a transmon qubit at 20$\,$mK and find a nonlinear dependence on the applied microwave power. 
The transmon-resonator system was operated in the strong dispersive
 regime, where the ac Stark shift ($2\chi$) due to a single microwave photon present in the resonator was larger than the linewidth
  ($\Gamma$) of the qubit transition.
 When the resonator was coherently driven at $5.474325\,$GHz, the
  transition spectrum of the transmon at  $4.982\,$GHz revealed 
  well-resolved peaks, each corresponding to an individual photon
   number-state of the resonator. 
 From the relative peak-heights we obtain the occupancy of the 
 photon-states and the average photon-occupancy  $\bar{n}$ 
 of the resonator.
 We observed a nonlinear variation of $\bar{n}$ with the applied 
 drive power $P_{rf}$ 
  for $\bar{n} < 5$ and compare our results to numerical simulations 
  of the system-bath master equation in the steady state, as well 
  as to  a semi-classical model for the resonator that includes the 
  Jaynes-Cummings interaction between the transmon and the resonator.
   We find good quantitative agreement using both models and analysis
    reveals that the nonlinear behavior is principally due to shifts in the
     resonant frequency caused by a qubit-induced Jaynes-Cummings nonlinearity.

\end{abstract}

\maketitle




\section{Introduction}

Many interesting quantum effects in superconducting circuits are
 fundamentally due to the nonlinearity of the Josephson junction. 
For a transmon \cite{Koch2007}, which is just a capacitively shunted
 Josephson junction with carefully chosen parameters, the junction causes sufficient anharmonicity in the energy eigenstates so that the two
  lowest levels can be isolated as a qubit. 
When this artificial `atom' couples strongly to the electromagnetic field of
 a resonator \cite{Wallraff2004,Wallraff2007} the same nonlinearity
  results in a variety of circuit QED (cQED) effects including photon
   number-splitting \cite{Gambetta2006, Schuster2007, Suri2013},
    nonclassical photonic states \cite{ Houck2007, Hoi2012,
     Kirchmair2013,Vlastakis2013} and the Autler-Townes effect
      \cite{Suri2013, Hoi2011, Novikov2013}.

Understanding of the nonlinear effects produced by the Josephson
 junction has also enabled improved qubit read-out \cite{Siddiqi2004,
 Reed2010} and the development of quantum-limited parametric
 amplifiers \cite{Bergeal2010,Hartridge2011}.
 The use of qubit readout techniques based on cQED effects is now
  wide-spread in quantum computing research. While most experimental
   efforts have been focused on using the techniques to examine qubits,
   the nonlinear quantum effects underlying such techniques are also of
 considerable interest. 

In this article, we examine nonlinear effects in the simplest cQED
 system, a resonator that is strongly coupled to a transmon. 
In particular, we study the nonlinear dependence of the average stored
 photon-number $\bar{n}$ in the resonator, versus the applied microwave
  power $P_{rf}$.
 When the coupling between the transmon and the resonator is large
  enough, the ac-Stark shift due to a single photon stored in the cavity
   can shift the qubit frequency by more than a spectral line-width
    resulting in ``photon number-splitting'' of the qubit spectrum 
     \cite{Schuster2007,Suri2013}. The photon occupancy $\bar{n}$ in the
 resonator can be directly found by measuring the transmon spectrum.  
 Gambetta \emph{et al.} \cite{Gambetta2006} predicted a nonlinear
  relationship between $\bar{n}$ and $P_{rf}$ due to the 
  Jaynes-Cummings interaction between the qubit and the resonator. 
Here we provide a detailed experimental examination of this 
effect in the small $\bar{n}$ limit. 
We also compare our results to numerical simulations of the 
system-bath master equation in the steady state, as well as  a 
semi-classical model for the resonator that includes the 
Jaynes-Cummings interaction between the transmon and
 the resonator, and find good quantitative agreement using both models.

In the following section, we first discuss the theory of the driven 
 transmon-resonator system using the Jaynes-Cummings Hamiltonian in
  the dispersive approximation and identify the leading nonlinear terms.
   We then set up a master equation to model losses and decoherence.
    In section \ref{sec:exp} we describe the experimental set-up used for
     the experiment and discuss the measured system parameters in
      section \ref{sec:exppar}. In section \ref{sec:datasim} we examine our
       results and compare them with the simulations, and then conclude in
        section \ref{sec:conclusion} with a summary and brief remarks on 
         nonlinear qubit read-out techniques.

   
 \section{Driven Jaynes-Cummings system}
 \label{sec:theory}
   Coupling of a transmon to the electromagnetic field in a microwave
    resonator can be modeled using a generalized Jaynes-Cummings
     Hamiltonian \cite{Jaynes1963, Boissonneault2010}:
\begin{widetext}
\begin {equation}
{H_{JC} = \hbar {\omega}_r a^{\dag} a  + \!\!\!\!\!\! \sum_{j = \{g,e,f...\}}\!\!\!\!\!\!\hbar{\omega}_j |j\rangle \langle j| +
 \!\!\!\!\!\! \sum_{j = \{g,e,f...\}}\!\!\!\!\!\! \hbar g_{j,j\!+\!1} \left\{a^{\dag} |j\rangle \langle j\!+\!1| + a |j\!+\!1\rangle \langle j| \right\}\,\,.}
\label{eqn:eqmlsjc}
\end{equation}
\end{widetext}
Here $\omega_r$ is the bare resonator  frequency, 
{$a^\dag\,$($a$) is the creation (annihilation) operator for the resonator mode, the} transmon states
{ $|j\rangle$ are labelled
\{${g,e,f,...}$\},
}
and $g_{j,j+1}$ is the  coupling strength
of the $|j\rangle \leftrightarrow |j+1\rangle$  transition 
of the transmon with the resonator mode.
In this form, the Josephson nonlinearity resides in the 
anharmonicity of the $\omega_j$.

\begin{figure}
\centering
\includegraphics[width=\columnwidth]{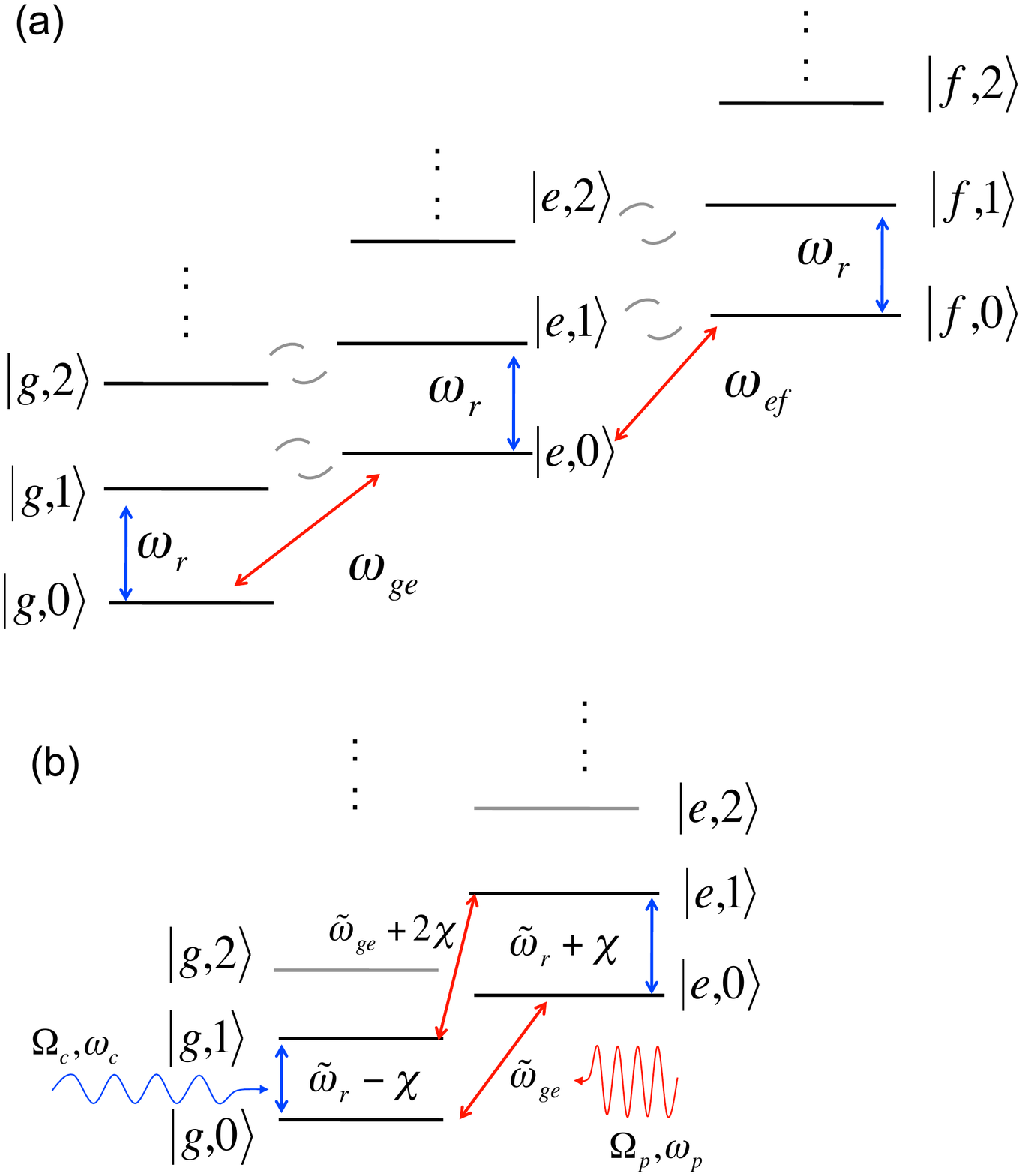}
\caption {(Color online) (a) Energy levels of the uncoupled 
transmon-resonator Jaynes-Cummings Hamiltonian for $m=3$
 transmon levels ($g, e, f$). (b) Energy levels after diagonalization
  in the dispersively coupled qubit approximation. The blue (red) 
  wavy arrow denotes coupler (probe) drive field with 
  strength $\Omega_{c(p)}$ and frequency $\omega_{c(p)}$.}
\label{fig:3jcladderdispersive}
\end{figure}

 The coupling term containing $g_{j,j+1}$ in Eq.~\ref{eqn:eqmlsjc} is the
  Jaynes-Cummings interaction (see Fig.~\ref{fig:3jcladderdispersive}(a))
   and it gives a block-diagonal structure to the Hamiltonian. 
   In particular, for an $m$-level system coupled to a resonator, 
   the $n$-excitation manifold is spanned by $m$ states $\{ |0,n\rangle,
    |1,n-1\rangle,\ldots, |m-1,n-m+1\rangle \}$. The corresponding
     $m\times m$ block matrix can then be diagonalized exactly for 
     small $m$ (numerically for large $m$) to compute the 
     eigenenergies of the coupled system.

The Hamiltonian can also be approximately
{diagonalized} in the  dispersive limit \cite{Blais2004,Carbonaro1979}
 corresponding to $\Delta_{j,j+1} \equiv \omega_{j,j+1}  - \omega_r \gg g_{j,j+1}$, 
 where $\omega_{j,j+1} \equiv \omega_{j+1}-\omega_j$
is the frequency of the $|j\rangle \leftrightarrow |j+1\rangle$ 
transmon transition. To diagonalize $H_{JC}$, we make 
a unitary transformation using:
\begin{widetext}
\begin{equation}
\mathcal{T} = \exp \left(\sum_{j = \{g, e, f \ldots\}} \lambda_{j,j+1} \left\{a |j+1\rangle \langle j| - a^\dagger |j\rangle \langle j +1 |\right\} \right),
\end{equation}
\end{widetext}
where $\lambda_{j,j+1} \equiv g_{j,j+1}/ \Delta_{j,j+1} \ll 1$ is 
the small parameter in which a perturbative expansion of 
the transformation can be carried out.  

Truncating the resulting transformed Hamiltonian to the two 
lowest transmon levels,  $H_{JC}$  can be written in the 
qubit approximation up to fourth order in $\lambda_{j,j+1}$ as  \cite{Boissonneault2010}:
 \begin{multline}
\mathcal{T}H_{JC}\mathcal{T}^\dagger \approx H_{JC}^{(4)} \approx \hbar \tilde{\omega}_r a^\dagger a + \hbar \frac{ \tilde{\omega}_{ge}}{2} \sigma_z \\ + \hbar \chi a^\dagger a \sigma_z + \hbar \zeta (a^\dagger a)^2 \sigma_z + \hbar \zeta' (a^\dagger a)^2,
\label{eqn:JCdisp4thorder}
\end{multline}
where $\sigma_z$ is the Pauli spin-matrix operating on 
the transmon $|g\rangle$ and $|e\rangle$ states,  $\tilde{\omega}_r \equiv \omega_r -\chi_{ef}/2$ is the Lamb-shifted resonator frequency,
 $\tilde{\omega}_{ge} \equiv \omega_{ge}+\chi_{ge}$ is the Lamb-shifted
  qubit frequency, and $\chi_{ge} = g_{ge}^2/ \Delta_{ge}$ and $\chi_{ef} = g_{ef}^2/\Delta_{ef}$ are the dispersive shifts of the resonator 
  due to the transmon $|g\rangle \to |e\rangle$ and $|e\rangle \to |f\rangle$ transitions respectively. 
In this approximation the total qubit state-dependent dispersive shift of the resonator frequency is given by $\chi \simeq \chi_{ge} - \chi_{ef}/2$ \cite{Blais2004}.
Thus Eq.~\ref{eqn:JCdisp4thorder}  includes perturbative shifts to the
 energy levels of the system due to the second-excited transmon 
 state $|f\rangle$ \cite{Bishop2010} although this state itself is 
 not explicitly included in Eq.~\ref{eqn:JCdisp4thorder}. 
We note that the anharmonicity was sufficiently large compared to
 the strength of the drive tones in our experiment that transitions to $|f\rangle$ were negligible.

 The term $\hbar\chi a^\dagger a \sigma_z$ in Eq.~\ref{eqn:JCdisp4thorder} represents an ac-Stark shift
  \cite{Blais2004} of the qubit transition frequency by an 
  amount $2\chi$ for every photon in the cavity 
  (see Fig.~\ref{fig:3jcladderdispersive}(b)). 
  We note that Eq.~\ref{eqn:JCdisp4thorder} also has two fourth 
  order terms that generate a Kerr-type nonlinearity. 
 The resonator-qubit cross-Kerr coefficient $\zeta$  and the 
 resonator self-Kerr coefficient $\zeta'$  \cite{Gambetta2006,
  Boissonneault2010}   are determined from:
 \begin{multline}
\zeta \approx \chi_{ef} \lambda^2_{ef} -2\chi_{ge}\lambda^2_{ge} \\+ \frac{7\chi_{ef}}{4}\lambda^2_{ge}   - \frac{5\chi_{ge}}{4}  \lambda^2_{ef}  \label{eqn:6zetaapprox}
\end{multline}
\begin{align}
\zeta' \approx (\chi_{ge}-\chi_{ef})(\lambda^2_{ge}+\lambda^2_{ef}). \label{eqn:6zetaprimeapprox}
\end{align} 

From Eq.~\ref{eqn:JCdisp4thorder}, the resonant frequency 
$\omega_r$ of the resonator can be seen to depend on the
 photon number $n = a^\dagger a$ according to
\begin{equation}
\omega_r (n) = \tilde{\omega}_r + \chi \sigma_z + \zeta n \sigma_z + \zeta' n.
\label{eqn:resfreqndep}
\end{equation}
We note that the $n$-dependence of the resonant frequency 
$\omega_r$ arises solely from the Kerr-type nonlinear terms up 
to this order of the approximate Hamiltonian. Thus the Kerr 
terms represent the lowest order approximation of the full 
nonlinearity arising from the Jaynes-Cummings interaction. 
It is this Kerr-type shift that causes the nonlinear dependence
 of the average photon-occupancy of the resonator $\bar{n}$ 
 discussed in section \ref{sec:datasim}.

In our experiment, we drive the transmon and the resonator using 
a ``probe'' tone and ``coupler'' tone respectively. 
We model the drives using the Hamiltonian
  \begin{multline}
H_d (t) = 
 - \frac{\hbar \Omega_c}{2}  (a e^{i\omega_c t}+a^\dagger e^{-i\omega_c t})  \\ - \frac{\hbar \Omega_p}{2} (\sigma^- e^{i\omega_p t}+\sigma^+ e^{-i\omega_p t}),  \label{eqn:3drivenjcham3}
\end{multline}
where $\omega_c$ ($\omega_p$) is the frequency of the 
coupler (probe) tone, $\Omega_c$ ($\Omega_p$) is the
 amplitude of the coupler (probe) tone driving the resonator 
 (qubit), and $\sigma^+$ and $\sigma^-$ are the raising and 
 lowering operators for the qubit. The Hamiltonian $H_d$ is 
 written in the rotating wave approximation in the limit of weak 
 driving ($\Omega_{c(p)} \ll \omega_{c(p)}$) \cite{Blais2004}. 
For simplicity, we also ignore terms of order $\lambda_{j,j+1}^2$ in $H_d$ that arise from the dispersive transformation $\mathcal{T}$.
 The driven Jaynes-Cummings Hamiltonian in the dispersive
  approximation can then be written as
\begin{align}
H \simeq H_{JC}^{(4)} + H_d (t)
\label{eqn:dispdrivenjc4thorder}
\end{align}

 To remove the explicit time-dependence in
  Eq.~\ref{eqn:dispdrivenjc4thorder} we transform into a frame
  rotating with the drives, using a unitary transformation
   \cite{Blais2004, Bishop2010} given by
\begin{equation}
U = e^{i \omega_c (a^\dagger a) t + i \omega_p \sigma_z t/2}.
\label{eqn:3rotatingframeofdrivesoperator}
\end{equation}
The transformed time-independent Hamiltonian can be written as
\begin{equation}
H_I  = U H U^\dagger + i \hbar \dot{U} U^\dagger
\end{equation}
which gives
\begin{multline}
H_I /\hbar =  \tilde{\Delta}_c a^\dagger a + \frac{\tilde{\Delta}_{p}}{2}\sigma_z +  \chi (a^\dagger a) \sigma_z \\ +  \zeta (a^\dagger a)^2 \sigma_z +  \zeta' (a^\dagger a)^2 \\- \frac{\Omega_c}{2} (a +a^\dagger) - \frac{\Omega_p}{2} (\sigma^-+ \sigma^+) \label{eqn:3drivenjctimeindepchap} 
\end{multline}
where $\tilde{\Delta}_c \equiv \omega_r - \omega_c$ and  $\tilde{\Delta}_p \equiv \tilde{\omega}_{ge} - \omega_p$. 

To include dissipation in the model we assume the
 transmon-resonator system is coupled to a thermal reservoir
  at temperature $T$. We write the master equation for the 
  system density matrix $\rho$ in the Born-Markovian approximation
   \cite{Blais2004, Thuneberg2013} as
\begin{multline}
\dot{\rho} = -\frac{i}{\hbar} [H, \rho] + \kappa_- \mathcal{D}[a]\rho + \kappa_+ \mathcal{D}[a^\dagger]\rho \\+ \Gamma_- \mathcal{D}[\sigma^-]\rho + \Gamma_+ \mathcal{D}[\sigma^+]\rho +\frac{\gamma_\varphi}{2}\mathcal{D}[\sigma_z]\rho  
\label{eqn:sme}
\end{multline}
where $\mathcal{D}[A_i]\rho \equiv A_i \rho A_i ^\dagger - \frac{1}{2}\left(A_i^\dagger A_i\rho + \rho A_i ^\dagger A_i\right)$, $H$ is 
given by Eq.~\ref{eqn:dispdrivenjc4thorder}, $\gamma_{\varphi}$ is 
the qubit dephasing-rate, $\kappa_-$ is the resonator photon 
loss-rate and $\kappa_+$ is the photon excitation-rate. 
Similarly, for the qubit, the relaxation rate is given by $\Gamma_-$ 
and the excitation rate by $\Gamma_+$. The loss-rate and 
excitation rate are related by a Boltzmann factor according 
to $\kappa_+ / \kappa_- \simeq e^{-\hbar \omega_r/k_B T}$ for 
the resonator, and $\Gamma_+ / \Gamma_- \simeq e^{-\hbar \omega_{ge}/k_B T}$ for the transmon \cite{Einstein1917}.
The explicit time-dependence in $H$ can be removed by 
transforming Eq.~\ref{eqn:sme} using the operator $U$
 in Eq.~\ref{eqn:3rotatingframeofdrivesoperator} to get
 \begin{multline}
\dot{\rho} = -\frac{i}{\hbar} [H_I, \rho] + \kappa_- \mathcal{D}[a]\rho + \kappa_+ \mathcal{D}[a^\dagger]\rho \\+ \Gamma_- \mathcal{D}[\sigma^-]\rho + \Gamma_+ \mathcal{D}[\sigma^+]\rho +\frac{\gamma_\varphi}{2}\mathcal{D}[\sigma_z]\rho.   
\label{eqn:6timeindependentmasterequation}
\end{multline}
 The steady state condition $\dot{\rho}=0$ reduces this master 
 equation to a set of coupled linear equations in the elements of
  the density matrix $\rho$. After truncating the number of photon
   levels, typically to around fifteen, the resulting equations were
    numerically solved \cite{Suri2015} to compare to the experimental
     results (see Sec.~\ref{sec:datasim}).  

\section{Experiment}
\label{sec:exp}
           
\begin{figure}
 \begin{center}
      { \includegraphics[width=\columnwidth ]{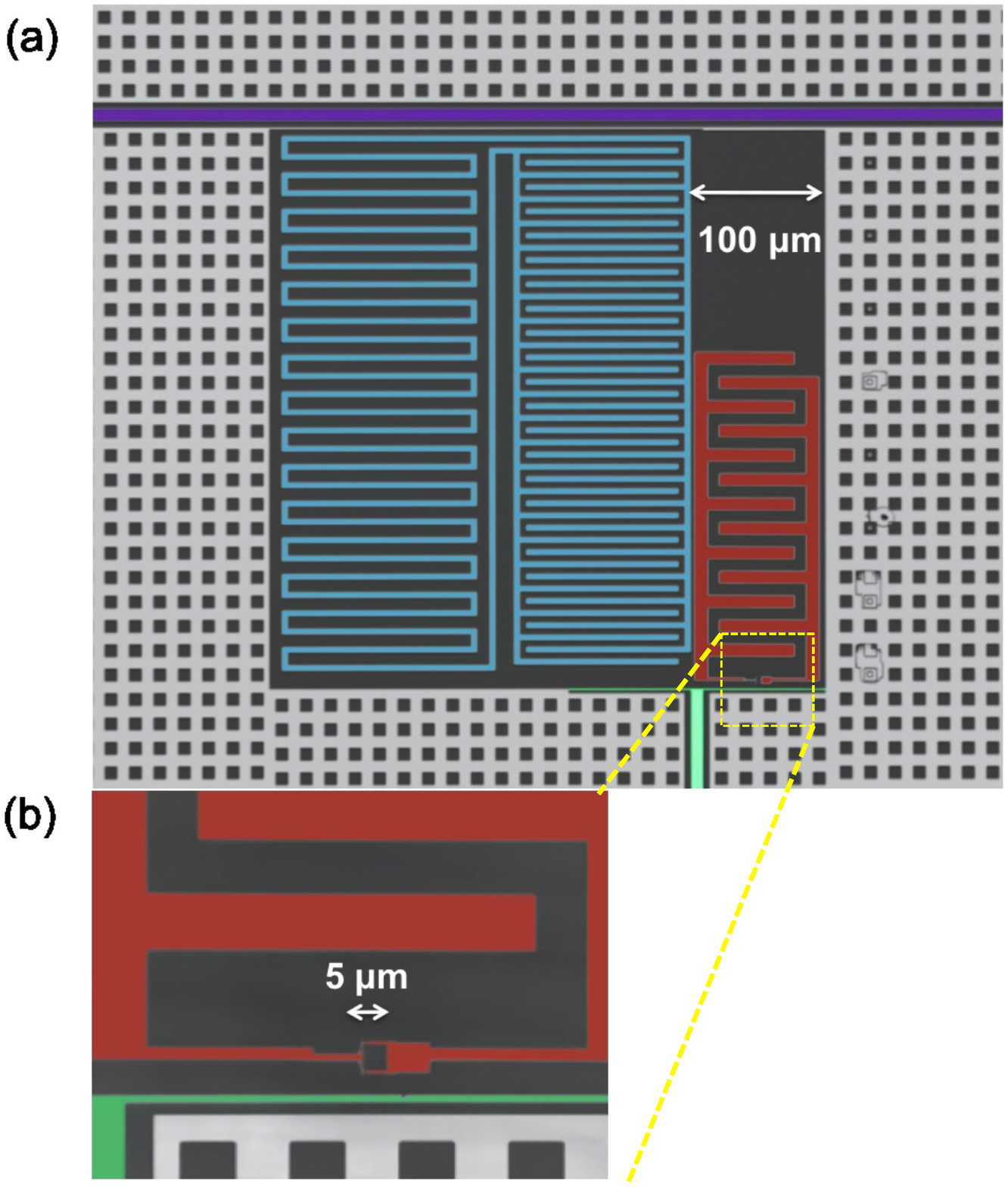}}
     \end{center}
     \caption{(Color online) (a) Colorized micrograph of the device \cite{Suri2015}.
       A lumped element resonator (blue) and transmon (red) are coupled
        to a coplanar waveguide transmission line (violet) and surrounded
         by a perforated ground plane (gray). The resonator consists of a
          meandering inductor and interdigitated capacitor. The transmon
           has two Josephson junctions in parallel to allow tuning of the
            transition frequency using an external magnetic field and on-chip
             flux bias line (green). (b) Detailed view of Josephson 
             junctions and flux bias line. }
            \label{fig:lev5zk9dev}
\end{figure}


Figure~\ref{fig:lev5zk9dev} shows a colorized micrograph of our
 transmon (red) coupled to a superconducting lumped-element resonator
  (blue)  \cite{Kim2011, Suri2015}. The device was patterned using
   photolithography and electron-beam lithography and was made 
   of thin-film aluminum on a sapphire substrate. 
The resonator is formed from a meandering inductor and an
 interdigitated capacitor  with a bare resonator frequency 
 $\omega_r/2\pi = 5.464\,$GHz. The resonator is coupled to an
  input/output coplanar waveguide transmission line (purple) which 
  is used to excite and measure the system. 

The transmon \cite{Koch2007} is formed from two Al/AlO$_x$/Al
 Josephson junctions that are shunted by an interdigitated capacitor
  with 13 fingers. Each finger has a width of 10$\,\mu$m, a length of 
  70$\,\mu$m, and is separated from adjacent fingers by 10$\,\mu$m,
   using e-beam lithography.
 The junctions, each of area $\approx 100 \times 100\,$nm$^2$, 
 were formed by double-angle evaporation \cite{Dolan1977, Suri2015}.
  They are connected to form a superconducting loop of 
 area $4 \times 4.5\, \mu$m$^2$ and the  loop is placed close to a
  shorted current bias line (green region in Fig.~\ref{fig:lev5zk9dev}). 
  This arrangement allows us to apply flux to the loop to finely tune the
   critical current of the parallel junctions and hence the transition
    frequency of the qubit. The normal-state resistance of the two
     Josephson junctions in parallel yielded a  maximum Josephson
      energy $E_{J,max}/h \approx 25\,$GHz and the transmon had a
       Coulomb charging energy  of  $E_{c}/h = 250\,$MHz. 
 This gave a maximum ground-to-first excited state transition 
 frequency $\omega_{ge,max}/2\pi \simeq (\sqrt{ 8E_{J,max} E_c} - E_c)/h = 7.1\,$GHz for the qubit.

The device was mounted in a  hermetically sealed copper box 
and attached to the mixing chamber of an Oxford Kelvinox 100 
dilution refrigerator with a base temperature of $20\,$ mK. 
To isolate the device from thermal noise at higher temperatures,
 the input microwave line to the device had a 10 dB attenuator 
 (Midwest microwave) mounted at 4 K, 20 dB at 0.7 K, and 30 dB 
  at 20 mK on the mixing chamber, for a nominal total attenuation
   of 60 dB. 
On the  output microwave line, two 18 dB isolators (Pantech) 
with bandwidths from 4 to 8 GHz were placed in series at 20 mK,
 followed by a 3 dB attenuator and a high electron mobility transistor
  (HEMT) amplifier (Caltech 4-8 GHz bandwidth, 40 dB gain) at 4 K. 

\begin{figure}
\centering
\includegraphics[width=\columnwidth]{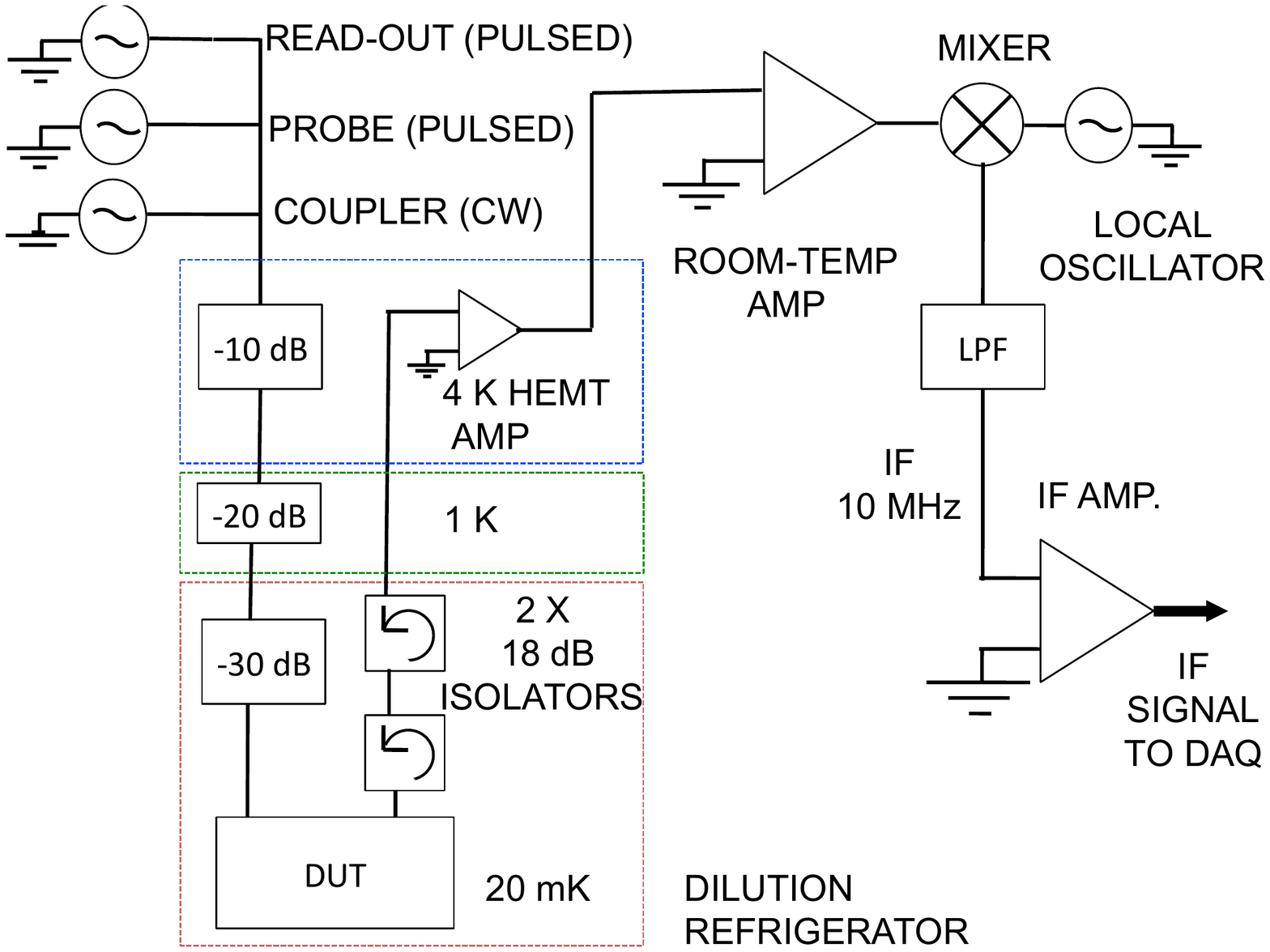}
\caption{(Color online) Schematic of the measurement set-up used for the number-splitting  measurements. }
\label{fig:atspecsetup}
\end{figure}

To observe the photon-number-split spectrum, three microwave 
tones were applied on the input line to the device: probe, coupler, 
and read-out (see Fig.~\ref{fig:atspecsetup}). 
The read-out and probe tones were pulsed on and off, while the 
coupler tone was applied continuously for the duration of the 
measurement.  
We used the high-power nonlinear Jaynes-Cummings read-out 
technique \cite{Reed2010, Boissonneault2010, Bishop2010} to
 measure the excited state probability of the transmon. 
   A probe pulse was first applied for a duration of $5\, \mu$s at 
   an amplitude just large enough to saturate the qubit transition 
   without causing large power broadening.
   We then waited $20\,$ns and applied a read-out pulse for 5$\,\mu$s 
   at a high power (\textit{i.e}. 50 dB larger than the power of the coupler
    tone) at the bare resonator frequency $\omega_r$. In our experiment,
     when the transmon is in the $|g\rangle$ state  a large transmissivity
      ($S_{21}$) of the read-out tone is observed and when the transmon
       is in an excited state ($|e\rangle$, $|f\rangle$ \emph{etc.}) a small
        transmissivity is observed.
    The output read-out microwave signal was amplified again 
    (Miteq 4-8 GHz, 30 dB gain) at room temperature and mixed 
    down to an intermediate frequency (IF) of 10 MHz 
    (see Fig.~\ref{fig:atspecsetup}).
     The IF signal along with a phase reference was digitized at 
     1 GSa/s \cite{Suri2015} using a data acquisition card and 
     the in-phase and quadrature components were demodulated
     before being recorded.


\section{System parameters}
\label{sec:exppar}
We determined the main system parameters from spectroscopic 
and time-domain measurements.
The resonator parameters were characterized by measuring 
the microwave transmission close to the resonant frequency 
as a function of input microwave power (data not shown). 
We measured a ``bare'' resonant frequency 
$\omega_r/2\pi = 5.464\,$GHz at large input powers.
 At small input powers, a ``dressed'' resonant frequency 
 of $(\tilde{\omega}_r - \chi)/2\pi=(\omega_r - \chi_{ge})/2\pi = 5.474325\,$GHz  
 was measured with the qubit in the ground state. 
 We also measured an internal quality factor $Q_I = 190,000$, 
 an external coupling quality factor $Q_e =  20,000$ and 
 a loaded quality factor $Q_L \equiv \omega_{r}/\kappa_- = 18,000$
  corresponding to a $\kappa_-= 2\pi (300)\,$kHz.
 
\begin{figure}
 \begin{center}
      { \includegraphics[trim=0cm 0cm 0 0, clip=true,width=\columnwidth]{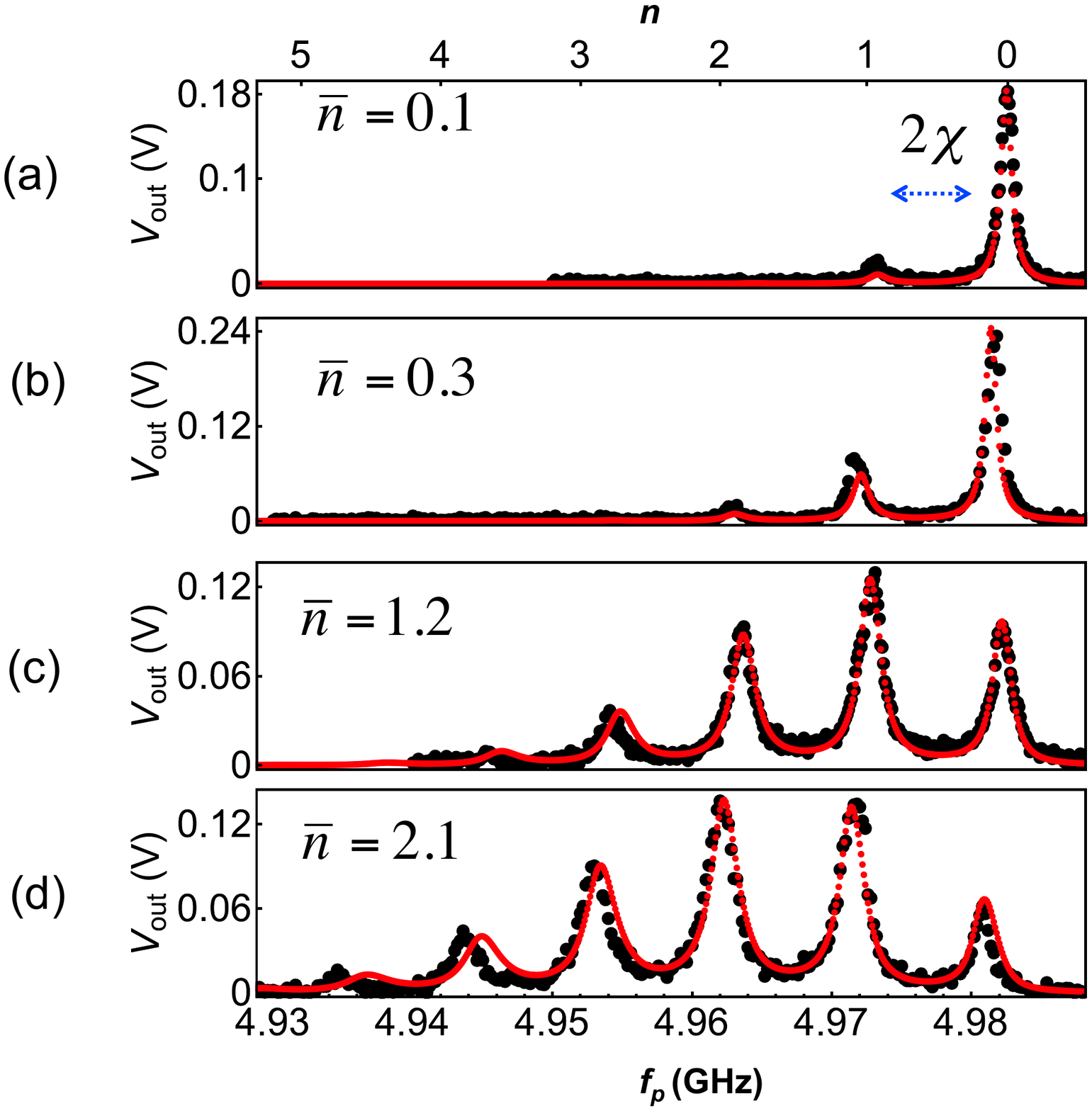}}
     \end{center}
     \caption{(Color online) Photon number-splitting in the transmon
      spectrum. Black is data. Red is steady-state solution of the 
      master equation Eq.~\ref{eqn:6timeindependentmasterequation}. 
      (a) Transmon spectrum with no coupler tone applied. 
      The primary qubit transition is seen at $\widetilde{\omega}_{ge}/2\pi = 4.982\,$GHz. 
      The peak  at $4.973\,$GHz is due to a residual population of $n_{th} =  0.1$ photons in the resonator.
       {(b)} Spectrum when a coupler tone was applied at 
       $\omega_c/2\pi = 5.474325\,$GHz and power  $P_{rf} = 2.5\,$aW, producing a population of  $\bar{n} = 0.3$ photons in the resonator. 
       {(c)} Coupler tone of power $P_{rf} =20 \,$aW,  $\bar{n} = 1.2$ photons in the resonator.
     {(d)} Coupler tone of power $P_{rf} =160 \,$aW,  $\bar{n} = 2.1$ photons in the resonator.}
     \label{fig:Numsplitsameaxis}
\end{figure}
The qubit transition frequency was tuned to $\widetilde{\omega}_{ge}/2\pi = 4.982\,$GHz 
using a combination of an external superconducting magnet and 
 the on-chip flux bias, corresponding to a detuning of 
 $\Delta_{ge} /2\pi= - 482\,$ MHz from the bare resonator frequency. 
The effective dispersive shift $\chi/2\pi = -4.65\,$MHz  was determined
 from the difference in the frequencies of the $n=0$ and $n=1$ 
 qubit spectral peaks (see Fig.~\ref{fig:Numsplitsameaxis}(a)).
From this we found $g_{ge}/2\pi = 70\,$MHz,  $\chi_{ge}/2\pi = -10.3\,$MHz, 
$\chi_{ef}/2\pi = -10.7\,$MHz and $g_{ef}/2\pi = 89\,$ MHz. 
The Kerr coefficients $\zeta/2\pi = 85\,$kHz and $\zeta' = -23\,$kHz 
were then calculated using Eq.~\ref{eqn:6zetaapprox} and Eq.~\ref{eqn:6zetaprimeapprox} respectively \cite{Suri2013,Suri2015}.  

Time-domain coherence measurements revealed a qubit relaxation time
 $T_{1} = 1/(\Gamma_- +\Gamma_{+})  = 1.6 \ \mu$s for the excited
  state of the qubit and a Rabi decay time  $T' = 1.6\ \mu$s \cite{Suri2015}.
 From these measurements, the pure dephasing rate was estimated at
  $\gamma_{\phi}  \equiv 1/T_\phi \approx 2\times10^5\,$s$^{-1}$.
  From qubit spectroscopy (see next section) an effective thermal bath
   temperature of $T = 120\,$mK was determined, even though the mixing
    chamber of the dilution refrigerator was at $20\,$mK.
     At this effective temperature,  
   $\kappa_+ / \kappa _- \simeq \Gamma_+ / \Gamma_- \simeq e^{-\hbar \omega_r / k_B T} \simeq 0.1$.
    

\section{Discussion of data and comparison with simulation}
\label{sec:datasim}
When the resonator is weakly driven using a continuous coupler tone at
 the dressed resonant frequency $\tilde{\omega}_r - \chi$, we expect this
  to yield a coherent state $|\alpha\rangle$ in the resonator such that 
  $| \alpha |^2 = \bar{n}$ in the steady state (see Appendix A). 
  The coherent state $|\alpha\rangle$ is a superposition of $|n\rangle$
   states with a well-defined photon number $n$ such that the probability
    $w_n$ of seeing $n$ photons obeys a Poisson distribution \cite{Glauber1963}.
  The qubit transition frequency is ac-Stark shifted (to lowest order in
   $\lambda$) by $\sim 2\chi n$ when the resonator is in the photon
    number-state $|n\rangle$. When $\chi \gg \Gamma$ (qubit line-width),
     individual peaks can be resolved in the qubit spectrum
      \cite{Schuster2007, Gambetta2006}, with the relative peak-heights
       obeying the Poisson distribution $w_n$. The average 
       photon-number $\bar{n}$ can then be calculated from the 
       qubit spectrum using a number-weighted average as
 \begin{equation}
\bar{n} = \frac{\sum_n  w_n n} {\sum_n w_n}.
\label{eqn:numphotweightedavg}
\end{equation}
   
Figure ~\ref{fig:Numsplitsameaxis}(a) shows  the transition spectrum 
of the transmon with no coupler tone applied to the resonator. 
The measured spectrum (black dots) shows a prominent peak at 
the qubit transition frequency of $\widetilde{\omega}_{ge}/2\pi = 4.982\,$GHz. 
The smaller spectroscopic peak, detuned  by -9.3 MHz at
  $(\widetilde{\omega}_{ge}+2\chi)/2\pi =4.973\,$GHz, is due to  
  one photon being present in the resonator. 
 Assuming that this was from thermal excitations, we can use the relative
  heights of the two spectral peaks and Eq.~\ref{eqn:numphotweightedavg} to estimate an 
  equilibrium thermal population of $n_{th} = 0.10$ photons,
   corresponding to a temperature of about $120\,$mK for the resonator.
    We note that qubit spectroscopy close to the $|e\rangle \to |f\rangle$
     transition frequency also revealed a small peak at $\omega_{ef}/2\pi$
      (data not shown) consistent with a thermal population of the qubit
       with an effective temperature of 120$\,$mK. This effective
        temperature is much higher than the $20\,$mK base temperature  of
         the dilution refrigerator, possibly due to leakage of infrared photons \cite{Corcoles2011} or 
 insufficient cooling of an attenuator stage on the input or output microwave lines. 

Figure \ref{fig:Numsplitsameaxis}(b) shows the measured transmon
 spectrum (black dots) when a weak coupler tone ($P_{rf} = 2.5\,$aW) is
  applied at  
  the resonant frequency $(\tilde{\omega}_r - \chi)/2\pi = 5.474325\,$GHz of the dressed resonator.
   Here, an increase in the height of the 
   $\widetilde{\omega}_{ge}+2\chi$ peak is observed and a spectral 
   peak at $\widetilde{\omega}_{ge}+4\chi$  appears. The 
   $n=0$ peak at $\widetilde{\omega}_{ge}$ is still the largest. 
   Figure \ref{fig:Numsplitsameaxis}(c) and (d) show the spectrum 
   for an applied resonator drive power of $20\,$aW and 160$\,$aW respectively. 
   In this case, five or six spectral peaks are clearly observed. 
   For $\bar{n} \gg n_{th} $, we found that the peak-heights obeyed a
    Poisson distribution \cite{Glauber1963, Schuster2007, Suri2013},
     while for $n \approx n_{th}$ we observed significant discrepancy.
 
 We also simulated the qubit spectrum using a steady-state solution of
  the master equation (Eq.~\ref{eqn:6timeindependentmasterequation}).
   The red curves in Fig.~\ref{fig:Numsplitsameaxis} are the steady-state
    solutions to the master equation with different amplitudes of the
     coupler tone corresponding to those of the  measured spectra (black).
      The parameters of the master equation were determined using
       independent spectroscopic and time-domain measurements as
        described in Sec.~\ref{sec:exppar}. Overall the simulation  
        agrees well with the data. We note that in
         Fig.~\ref{fig:Numsplitsameaxis}(d), the simulated spectrum
          deviates from data near the $n=4$ and $n=5$ photon peaks. 
          This discrepancy probably occurred because the simulation 
          did not include terms higher than  the fourth order Kerr-type
           terms (see Eq.~\ref{eqn:3drivenjctimeindepchap} and
            Eq.~\ref{eqn:6timeindependentmasterequation}).

\begin{figure}[tbp]
\centering     
\includegraphics[width=\columnwidth]{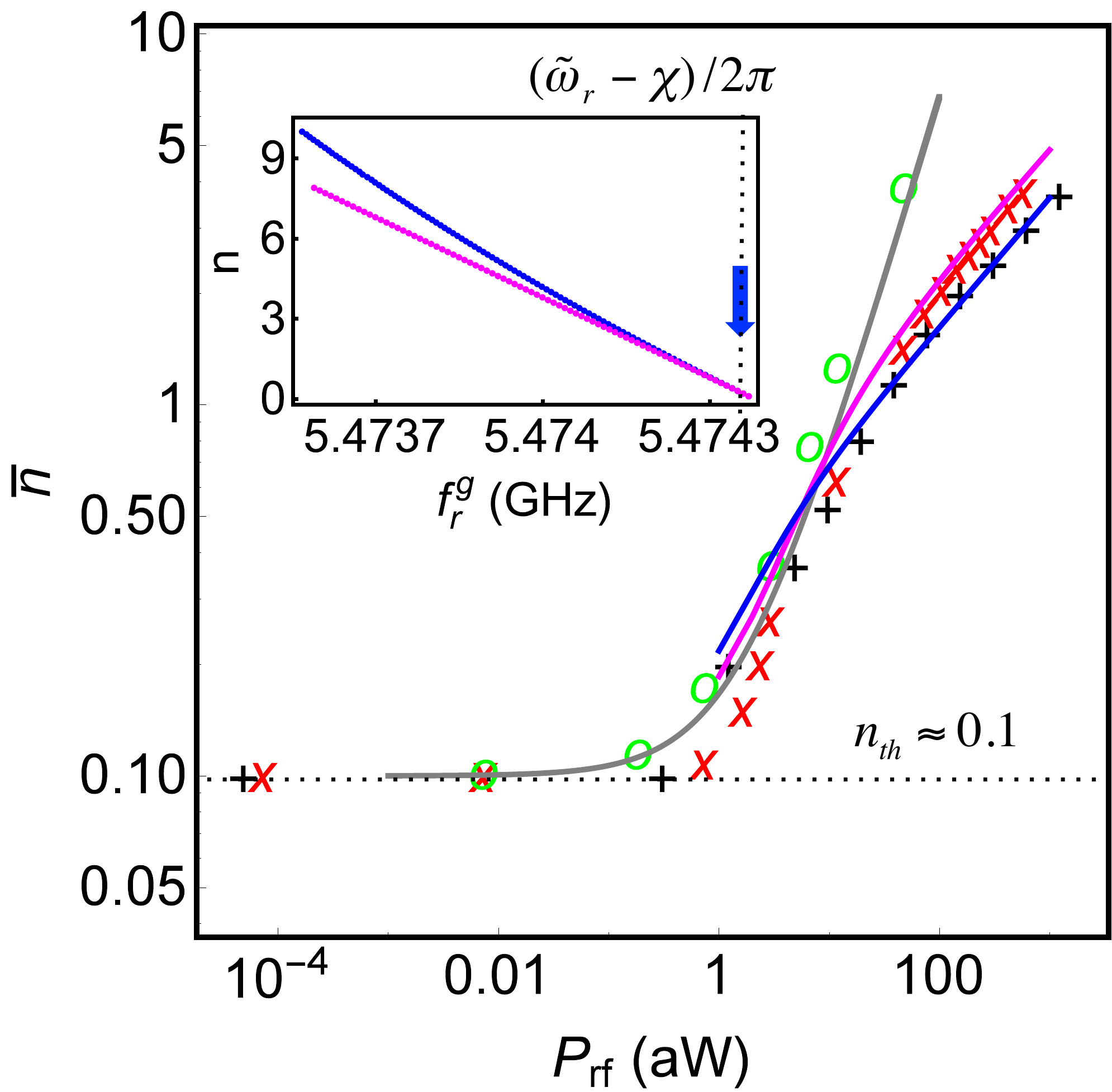}
 \caption{(Color online) Average photon number $\bar{n}$ in the
  resonator as a function of the applied microwave power $P_{rf}$.
   Black `+' marks show measured $\bar{n}$ from weighted average of the
    qubit spectral peaks. Gray curve is the classical linear model
     (Eq.~\ref{eqn:nbarlinearclassical}). Magenta curve is solution of 
     the semiclassical nonlinear equation (Eq.~\ref{eqn:nbarnonlinearkerr})
      where a Kerr-type resonant frequency shift is included. Blue curve is
       the solution of the nonlinear equation
        (Eq.~\ref{eqn:nbarnonlinearfull}) where higher order 
        nonlinearities are included through exact diagonalization.
         Red `x' marks show $\bar{n}$ computed  from a weighted
          average of the spectra simulated using the system Master
           equation (Eq.~\ref{eqn:6timeindependentmasterequation})
            including the Kerr-type nonlinearities (up to fourth order in
             $\lambda$). Green circles show $\bar{n}$ computed  from
              weighted average of spectra simulated using the system master
               equation (Eq.~\ref{eqn:6timeindependentmasterequation})
excluding Kerr nonlinearities.  Inset shows shift of resonant frequency
 when the qubit is in $|g\rangle$ with increasing photon number
  calculated by including Kerr terms (magenta curve) from
   Eq.~\ref{eqn:resfreqndep} and including higher-order nonlinearities
    through exact diagonalization of Eq.~\ref{eqn:eqmlsjc} (blue curve).}
     \label{fig:photonnumvspowerdatatheorysim}
\end{figure}

  The average number of photons $\bar{n}$ in the resonator was
   determined from each measured photon number-split spectrum using
    Eq.~\ref{eqn:numphotweightedavg} \cite{Suri2013}.
  Fig.~\ref{fig:photonnumvspowerdatatheorysim} shows the average
   number of photons $\bar{n}$ versus the applied power $P_{rf}$ in
    attowatts (black `+' marks). For very weak driving $P_{rf} < 1\,$aW, the
     thermal photon population $n_{th} = 0.1$ is the dominant contribution to $\bar{n}$. 
  Above an applied power of  $P_{rf} > 1\,$aW, $\bar{n}$  increases
   monotonically, but nonlinearly.

To understand this nonlinear dependence of $\bar{n}$ on $P_{rf}$, we
 first consider using a semi-classical approach. 
 Classically, we expect the mean occupancy of photons in the resonator
  to be given by  
 \begin{equation}
\bar{n} = 0.1+ \frac{P_{rf} / 4\hbar Q_e}{\delta^2 + (\kappa_-/2)^2},
\label{eqn:nbarnonlineargeneral}
\end{equation}
where $\delta = \omega_c - (\tilde{\omega}_{r} - \chi)$ is the detuning 
of the coupler drive from the resonant frequency 
($\tilde{\omega}_r - \chi)$ \cite{Gambetta2006, Suri2013, Suri2015}. 
The additional 0.1 accounts in an \emph{ad hoc} way for the 
equilibrium thermal photon population $n_{th}$.
In general, when $\delta$ is constant, the second term
 in Eq.~\ref{eqn:nbarnonlineargeneral} gives a linear relation
  between $\bar{n}$ and $P_{rf}$. On resonance, $\delta = 0$ 
  and  Eq.~\ref{eqn:nbarnonlineargeneral} reduces to 
\begin{equation}
\bar{n} = 0.1+ (Q_L/Q_e)P_{rf}/\kappa_- \hbar \omega_r.
\label{eqn:nbarlinearclassical}
\end{equation}

  The gray curve in the Fig.~\ref{fig:photonnumvspowerdatatheorysim}
   shows the linear model given by Eq.~\ref{eqn:nbarlinearclassical}.  
   The experimental data deviates markedly from the linear model even 
   at an average occupancy of one photon in the resonator. Note that
    the critical number of photons \cite{Blais2004} for 
    $g_{ge}/2\pi = 70\,$MHz and $\Delta_{ge}/2\pi = -482\,$MHz
     is $n_{crit} = \Delta_{ge}^2/4g_{ge}^2 \simeq 12$. 
   Therefore, the deviation from linearity happens at 
   $\bar{n}\ll n_{crit}$, as predicted by Gambetta \emph{et al.} \cite{Gambetta2006}.
  
  Although Eq.~\ref{eqn:nbarlinearclassical} is not well-obeyed over 
  the full range of $\bar{n}$,  the measured mean photon occupancy
   is nearly linear for $0.1 < \bar{n} < 0.5$. Using 
    Eq.~\ref{eqn:nbarlinearclassical} and the measured $\bar{n}$ from
     spectrum in Fig.~\ref{fig:Numsplitsameaxis}(b), we find an 
     attenuation of  $\alpha = 65\,$dB for the input microwave line. 
     This estimate agrees reasonably well with the nominal attenuation 
     of $60\,$dB (see Fig.~\ref{fig:atspecsetup}). 
  
   The nonlinearity can be captured by realizing that $\delta(\bar{n})$ is
    in general a function of $\bar{n}$.   As a first approximation, we
     include the shift in the resonant frequency only up to the 
     Kerr-type terms (Eq.~\ref{eqn:resfreqndep}). This gives rise 
     to the nonlinear equation
   \begin{equation}
\bar{n} =0.1+ \frac{P_{rf} / 4\hbar Q_e}{\{(\zeta' -\zeta)\bar{n}\}^2  + (\kappa_-/2)^2},
\label{eqn:nbarnonlinearkerr}
\end{equation}
   where the lowest order approximation 
   $\delta = \omega_c - (\tilde{\omega}_r - \chi) \approx (\zeta' - \zeta) \bar{n}$ 
   from Eq.~\ref{eqn:resfreqndep} is used. The Kerr-type 
   coefficients $\zeta /2\pi = 85\,$kHz and $\zeta'/2\pi = -23\,$kHz 
   were estimated using Eq.~\ref{eqn:6zetaapprox} 
   and Eq.~\ref{eqn:6zetaprimeapprox}. 
   All of the parameters in Eq.~\ref{eqn:nbarnonlinearkerr} 
   were determined from spectroscopy, allowing us to solve it in 
   a self-consistent manner for $\bar{n}$ with no fit parameters. 
    
   The magenta curve in Fig.~\ref{fig:photonnumvspowerdatatheorysim}
    shows the solution of Eq.~\ref{eqn:nbarnonlinearkerr}.  We find 
    that this qualitatively captures the nonlinear variation of $\bar{n}$ 
    with $P_{rf}$, and suggests that it arises due to the Kerr terms, 
    which cause a change in the resonant frequency as $\bar{n}$
     is varied. The deviation from data for $\bar{n}>0.5$ is possibly 
     caused by ignoring terms of order higher than the Kerr terms.

      We extended Eq.~\ref{eqn:nbarnonlinearkerr} to include higher 
      order nonlinearities by writing
       \begin{equation}
\bar{n} =0.1+ \frac{P_{rf} / 4\hbar Q_e}{[\delta(\bar{n})]^2  + (\kappa_-/2)^2},
\label{eqn:nbarnonlinearfull}
\end{equation}
where the detuning $\delta(n) = \omega_c - \omega_r^g (n)$ 
was calculated from exact diagonalization of the generalized 
Jaynes-Cummings Hamiltonian (Eq.~\ref{eqn:eqmlsjc}) with 
10 transmon levels in the calculation. 
Equation~\ref{eqn:nbarnonlinearfull} can be solved self-consistently 
to find the mean occupancy $\bar{n}$ as a function of $P_{rf}$. 

The solution of Eq.~\ref{eqn:nbarnonlinearfull} is plotted as the 
blue curve in Fig.~\ref{fig:photonnumvspowerdatatheorysim}. We
 find that this agrees closely with the data, particularly in the 
 $\bar{n} > 0.5$ range. This  semiclassical model  deviates
  from the data in the range $0.1 < \bar{n} < 0.5$, possibly 
  due to the \emph{ad hoc} nature in which the thermal 
  photon population was included in Eq.~\ref{eqn:nbarnonlinearfull}.   

The blue curve in the inset to
 Fig.~\ref{fig:photonnumvspowerdatatheorysim} shows the 
calculated frequency of the resonator when the qubit is in the 
ground state ($f_r^g = \omega_r^g/2\pi$)  as  photon-number $n$ 
in the resonator is varied, using an exact diagonalization of the
 generalized Jaynes-Cummings Hamiltonian (Eq.~\ref{eqn:eqmlsjc}). 
 The ``bare'' parameters of the Hamiltonian   were determined 
 from spectroscopic measurements (see Sec.~\ref{sec:exp}) 
 and for the calculation, we used up to 10 transmon levels for 
 greater accuracy. The magenta curve in the inset 
 to Fig.~\ref{fig:photonnumvspowerdatatheorysim} 
 corresponds to the variation of $f_r^g$ with $n$ calculated 
 using just the Kerr nonlinearity (Eq.~\ref{eqn:resfreqndep}).

We note that in the experiment, the coupler frequency is fixed 
at $\omega_c = \tilde{\omega}_r - \chi = 2\pi (5.474325)\,$GHz 
(indicated by the blue arrow in the inset). The good agreement 
between the blue curve (solution to Eq.~\ref{eqn:nbarnonlinearfull}) 
and the data in Fig.~\ref{fig:photonnumvspowerdatatheorysim}  
indicates that it is the change in the resonant frequency with 
$\bar{n}$ that causes the nonlinear power-dependence of 
$\bar{n}$. Conversely, if the  coupler frequency were adjusted to 
always drive on resonance, we expect $\bar{n}$ would agree with 
the classical linear model (gray curve in Fig.~\ref{fig:photonnumvspowerdatatheorysim}).

 For comparison, we also simulated the behaviour of the system using
  the full master equation (Eq.~\ref{eqn:6timeindependentmasterequation}). The parameters
   used in the simulation were measured experimentally as discussed
    in Sec.~\ref{sec:exppar}.
    The red `x' marks in Fig.~\ref{fig:photonnumvspowerdatatheorysim}
     represent the average photon number computed by applying
      Eq.~\ref{eqn:numphotweightedavg} to spectra numerically simulated
       using the master equation (Eq.~\ref{eqn:6timeindependentmasterequation}) where the 
       system Hamiltonian included terms up to the Kerr-type nonlinearities.
    As expected, $\bar{n}$  calculated from the master equation 
    solution shows a nonlinear variation of $\bar{n}$ with $P_{rf}$ 
    and it agrees closely with the semiclassical calculation which 
    included up to the Kerr terms (magenta curve). We note that 
    the $\bar{n}$ from the master equation simulation shows a small 
    but consistent deviation from the data (black `+' marks) for 
    $\bar{n} > 1$. This disagreement at higher power is likely 
    due to contribution from higher order nonlinearities which 
    have been ignored. 
    
    To verify that the Kerr terms cause the nonlinearity in $\bar{n}$, 
    we set the Kerr terms to zero in the master equation simulation. 
    The green circles in Fig.~\ref{fig:photonnumvspowerdatatheorysim}
     show $\bar{n}$ from this simulation, and the resulting curve falls 
     near to the classical linear theory (gray curve)as $\bar{n}$ increases.
    
     Finally we note that the $\bar{n}$ from master equation simulations
      (red `x' marks and green circles) follow the data in the range of $0.1 < \bar{n} < 0.5$ 
      more closely than the semiclassical simulations (blue and magenta).     
We believe this is because in our semiclassical analysis the thermal
 population ${n_{th}} = 0.1$ is included in a completely \emph{ad hoc}
  manner. No such approximation was made in the master equation
   simulations.
In general, at any non-zero temperature 
($\kappa_+ \neq 0, \Gamma_+ \neq 0$), the probability 
distribution $w_n$ is not Poissonian \cite{Schuster2007, Thuneberg2013}.  
A semiclassical model that better fits the data in the range
 $0.1<\bar{n}<0.5$, encompassing both coherent and thermal 
 population of photons, is beyond the scope of this paper.

  \section{Conclusion}
  \label{sec:conclusion}
  In conclusion, we have measured a transmon coupled to a 
  lumped-element resonator and observed a nonlinear dependence 
  of the photon occupancy of the resonator $\bar{n}$ versus the 
  applied microwave power $P_{rf}$. 
  The nonlinearity sets in at $\bar{n} \approx 1$, an order of
   magnitude smaller than the critical photon-number 
   $n_{crit} \approx 12$ that determines the validity of the dispersive
    approximation. 
  We found that the nonlinear dependence of $\bar{n}$ on $P_{rf}$ 
  is caused by  the fourth order (Kerr-type) terms in the system
   Hamiltonian. 
  We compared our experimental results to numerical simulations 
  of the system-bath master equation in the steady state, as well 
  as semi-classical models for a driven damped resonator and found
   good agreement when the variation of the resonant frequency with
    $n$ is included.
  
  Finally, we note that the high-fidelity read-out of the transmon state
   used in our experiment is based on the Jaynes-Cummings 
   nonlinearity \cite{Reed2010, Bishop2010, Boissonneault2010}. 
   While this read-out allows single-shot measurements of the 
   transmon state \cite{Reed2010, Paik2011, Chow2012},  the 
   read-out fidelity ($\sim 80\%$) we typically achieve is much 
   less than the fidelity of state preparation ($\sim 98\%$)  \cite{Reed2010}. The system master equation may be used to
    model this technique and quantitative understanding of the
     nonlinearities may lead to new or improved read-out techniques. 
     For example, since this read-out depends on the strength of
      the Jaynes-Cummings interaction, the relative detuning between 
      the qubit and resonator frequencies could play a role in the 
      read-out fidelity \cite{Boissonneault2010}and study of this
       dependence may yield improvements in the read-out fidelity. 
  
  \section*{Acknowledgements}
  The authors would like to thank Rusko Ruskov for useful discussions.
   F. C. W. would like to acknowledge support from the Joint Quantum
    Institute and the State of Maryland through the Center for
     Nanophysics and Advanced Materials. 
  \appendix
  \section{Coherent states as the solution of the driven damped oscillator}
  Here we review  the steady state solutions of the master equation 
  for a driven damped harmonic oscillator  \cite{Glauber1963}.  
  Ignoring the transmon, the master equation for a driven dissipative oscillator at $T=0\,$K  is 
\begin{equation}
\dot{\rho} = -\frac{i}{\hbar} [H_{osc} , \rho ] +\kappa_- \mathcal{D}[a]\rho
\label{eqn:smeosc}
\end{equation}
where
\begin{equation}
H_{osc} = \hbar{\omega_r} a^\dagger a -\frac{\hbar \Omega_c}{2}(a e^{i\omega_c t} + a^\dagger e^{-i\omega_c t})
\end{equation}
and  $\omega_r$ is the resonant frequency of the oscillator. 
The following equation can then be obtained from Eq.~\ref{eqn:smeosc}:
\begin{multline}
\frac{\mathrm{d}\langle a^\dagger a \rangle}{\mathrm{d}t} =\frac{\mathrm{d} \mathrm{Tr} [\rho . a^\dagger a ]}{\mathrm{d}t} =-\kappa_- \langle a^\dagger a \rangle \\+ \frac{i \Omega_c}{2} \left( \langle a^\dagger \rangle e^{-i \omega_c t} - \langle a \rangle e^{i \omega_c t} \right).
\label{eqn:numphotdecayclassical}
\end{multline}
 Equation \ref{eqn:numphotdecayclassical} confirms that photons 
 decay from the resonator at a rate set by $\kappa_-$, in the absence 
 of driving, \emph{i.e.} $\Omega_c  =0$, and the resonator reaches 
 the vacuum state $|0\rangle$ in steady state.

To show that the Glauber coherent states \cite{Glauber1963} are
 the steady state solutions of the master equation for the driven
  resonator, we first start with the Hamiltonian of the driven resonator:
\begin{align}
H &= \hbar \omega_r a^\dagger a -\frac{\hbar \Omega_c}{2} \left ( a^\dagger e^{-i\omega_c t} + a e^{i \omega_c t}\right).
\label{eqn:drivenoscham}
\end{align}

We then transform the Hamiltonian in Eq.~\ref{eqn:drivenoscham} 
using the Glauber displacement operator $D^\dagger(\alpha)$ where 
\begin{equation}
D(\alpha) = e^{\alpha a^\dagger  - \alpha^* a},
\end{equation}
where $\alpha$ is  a complex number characterizing the coherent state. 
Using
\begin{equation}
D^\dagger(\alpha)a^\dagger D(\alpha) = a^\dagger + \alpha^*,
\label{eqn:3displacementopdefstar}
\end{equation}
the transformed Hamiltonian becomes:
\begin{multline}
\tilde{H} = D^\dagger (\alpha) H D(\alpha) + i\hbar \frac{\partial D^\dagger (\alpha)}{\partial t} D(\alpha)\\
= \hbar \omega_r a^\dagger a + \hbar a^\dagger [\omega_r \alpha -  \frac{\Omega_c}{2}e^{-i\omega_c t} - i \dot{\alpha}] \\+ \hbar a [\omega_r \alpha^*- \frac{\Omega_c}{2}e^{i\omega_c t} + i \dot{\alpha}^*] 
\label{eqn:displaceddrivenhamosc}
\end{multline}

The master equation Eq.~\ref{eqn:smeosc} can also be transformed by $D^\dagger (\alpha)$ to read:
\begin{multline}
\dot{\rho}=-\frac{i}{\hbar} [\tilde{H},\rho] + \kappa_- \mathcal{D}[a] \rho  \\+ \frac{\kappa_-}{2} (\alpha^* [a,\rho] - \alpha[a^\dagger, \rho]) 
\end{multline}

{Expanding $\tilde{H}$ from Eq.~\ref{eqn:displaceddrivenhamosc}, we have}
\begin{multline}
\dot{\rho} = -i \omega_r [a^\dagger a, \rho] + \kappa_- \mathcal{D}[a]\rho \\-i [a^\dagger,\rho]\left(\omega_r \alpha - \frac{\Omega_c}{2} e^{-i \omega_c t} - i \dot{\alpha} -i \frac{\kappa_-}{2} \alpha\right) \\
  -i [a,\rho]\left(\omega_r \alpha^* - \frac{\Omega_c}{2} e^{i \omega_c t} +i \dot{\alpha}^* +i \frac{\kappa_-}{2} \alpha^*\right).
 \label{eqn:3rhodotdisplaced}
\end{multline}

We are free to choose $\alpha$ such that the terms in the parentheses
 in Eq.~\ref{eqn:3rhodotdisplaced} are zero, that is: 
\begin{align}
\omega_r \alpha - \frac{\Omega_c}{2} e^{-i \omega_c t} - i \dot{\alpha} -i \frac{\kappa_-}{2} \alpha = 0 \label{eqn:3drivendampedharmoscalpha}\\
\omega_r \alpha^* - \frac{\Omega_c}{2} e^{i \omega_c t} +i \dot{\alpha}^* +i \frac{\kappa_-}{2} \alpha^* = 0
\label{eqn:3drivendampedharmoscalphastar}
\end{align}
 Equations \ref{eqn:3drivendampedharmoscalpha}  and
  \ref{eqn:3drivendampedharmoscalphastar} are the same as the
   equations of motion of a classical oscillator \cite{Thuneberg2013}. With this choice of $\alpha$, the transformed master equation
    Eq.~\ref{eqn:3rhodotdisplaced} becomes simply
\begin{align}
\dot{\rho} = -i \omega_r [a^\dagger a, \rho]& + \kappa_- \mathcal{D}[a]\rho \label{eqn:3rhodotosc}
\end{align}
which is nothing but the master equation for an undriven oscillator,
 \emph{i.e.} Eq.~\ref{eqn:smeosc} with $\Omega_c = 0$. 

We already noted that the steady state solution for the undriven
 oscillator is the ground state or vacuum state $|0\rangle$. 
 However to arrive at Eq.~\ref{eqn:3rhodotosc} we began by
  transforming the Hamiltonian $H$ using the displacement operator
   $D^\dagger (\alpha)$.  If the untransformed steady state solution is 
   $|\psi_{ss}\rangle$, then we can write
\begin{equation}
D^\dagger (\alpha) |\psi_{ss}\rangle = |0\rangle
\label{eqn:3steadystatesoldisplaced}
\end{equation}
Noting that $D^\dagger (\alpha) = D (-\alpha)$, we find
\begin{align}
 |\psi_{ss}\rangle = D(\alpha) |0\rangle = |\alpha \rangle, 
\end{align}
where $|\alpha\rangle$ is a Glauber coherent state \cite{Glauber1963}.
Thus a damped oscillator at $T=0\,$K that is initially in the ground state
 $|0\rangle$ goes into a coherent state $|\alpha\rangle$ under harmonic
  driving,
where $\alpha$ satisfies the classical equation of motion
 Eq.~\ref{eqn:3drivendampedharmoscalpha}. 
Note that at a non-zero temperature, the distribution of 
Fock-states becomes thermal, while  $\alpha$ satisfies the classical
 oscillator equations \cite{Thuneberg2013}.

 Considering the coupled transmon-resonator system, we note that 
 up to second order in the dispersive approximation 
 ($\chi \sigma_z (a^\dagger a)$), the only effect the qubit has on 
 the resonator is to shift the resonant frequency. In this case the 
 above analysis can be applied, since the resonator is still linear. 
 We note that in the presence of higher-order nonlinearities, the 
 Glauber displacement transformation of the master equation is 
 non-trivial.



\end{document}